\documentstyle[twoside,fleqn,espcrc2,psfig]{article}

\newcommand{\AmS}{{\protect\the\textfont2
  A\kern-.1667em\lower.5ex\hbox{M}\kern-.125emS}}
\def\lsim{\raise0.3ex\hbox{$\;<$\kern-0.75em\raise-1.1ex\hbox{$\sim\;$}}}
\def\gsim{\raise0.3ex\hbox{$\;>$\kern-0.75em\raise-1.1ex\hbox{$\sim\;$}}}

\title{ {\vskip -4.0cm 
\hfill{\small hep-ph/9712455}\\
\vskip -0.2cm 
\hfill{\small FTUV/97-71}
\vskip -0.2cm 
\hfill{\small IFIC/97-102}
\vskip -0.25cm 
\hfill{\small FISIST/14-97/CFIF}}
\vskip 1.1cm
Borexino as a test of solar matter 
density fluctuations
\thanks{Talk given by H. Nunokawa
in Taup97, Gran Sasso, Italy, September, 1997}
}

\author{H. Nunokawa\address{Instituto de F\'{\i}sica Corpuscular - C.S.I.C., 
Departament de F\'{\i}sica Te\`orica, Universitat de Val\`encia\\
46100 Burjassot, Val\`encia, Spain},
A. Rossi\address{Dept. de Fisica, Inst. 
Superior Tecnico, 1096 Lisbon Codex, Portugal},
V. Semikoz$^{a,}$\address{Institute of the Terrestrial Magnetism, 
the Ionosphere and 
Radio Wave Propagation of the Russian Academy of Sciences, 
IZMIRAN, Troitsk, Moscow region, 142092, Russia}
and  
J.W.F.Valle$^a$
}
       
\begin{document}

\begin{abstract}
This talk summarizes some results of our recent work
focusing on the possibility to test solar matter 
density fluctuations by the future Borexino experiment. 
\end{abstract}

\maketitle

\section{Introduction}
The solar neutrino deficit observed in Homestake, 
SAGE, GALLEX and Kamiokande experiments \cite{solar} 
has been confirmed by the new result from 
Super-Kamiokande experiment \cite{SK}. 
The most successful explanation of the deficit is given by 
the resonant flavor conversion of neutrinos (MSW mechanism) 
\cite{MSW} in the solar interior. 

Based on the result obtained in ref. \cite{noise}, 
where we have studied the impact of random density perturbation
on the MSW solution to the solar neutrino problem, 
this talk features the possibility to test such 
fluctuations by the future Borexino experiment. 
The effect of density perturbations on the neutrino 
propagation has also been studied in refs. 
\cite{noise0,BL,torrente,burgess1,burgess2}. 

\section{Basic assumptions and general features of the effect}
In general, the matter density in the solar interior may 
fluctuate around its mean values as, 
\begin{equation}
\rho = \langle \rho \rangle+ \delta \rho,
\end{equation}
where the symbol $\langle...\rangle$ means to take average. 
We assume that the perturbation is random
and have a Gaussian distribution according to the following
correlation function:
\begin{equation}
\langle \delta \rho(r)\rho(r')\rangle \sim 
\xi^2 \langle \rho\rangle^2
L_0 \delta(r-r')
\end{equation}
where $\xi \equiv \delta \rho /\rho$ is the magnitude of 
the fluctuation, $L_0$ is the correlation length, which 
corresponds to the scale of the fluctuation. 

Here we only give the basic idea of how we can compute the survival 
probabilities of neutrinos in the presence of fluctuations 
and refer the reader to ref. \cite{noise} for detail. 
We start from the system of usual two-flavor evolution equations 
for neutrinos in matter but with the fluctuation term in eq. (1). 
The system of equations  is averaged over 
the fluctuation ensemble, taking into account the relation (2), 
{\it before} computing the final probability.  
In this way the evolution equations for the averaged survival 
probability $\langle P \rangle$ 
already include the effect of the fluctuation. 
An alternative approach, taking the ensemble average of 
$P$ {\it after} solving the equation for many different 
profiles of $\delta \rho$ is considered in ref. 
\cite{burgess2}. 

An important assumption in our analysis is
\begin{equation}
\label{correlation}
l_{free} \ll L_0 \ll \lambda_m, 
\end{equation}
where $l_{free}\sim 10 $ cm is the mean free path of the electrons
in the solar medium and $\lambda_m$ is the neutrino oscillation length 
in matter. 
We show the effect of density fluctuation on the MSW 
mechanism in Fig. 1. 
As we can see from the figure the general feature of the
effect 
is to suppress the MSW mechanism. 
Although the effect seems to be rather large, the two MSW 
solutions (large and small mixing angle) are found to be 
rather stable \cite{noise}, which 
implies that the current solar neutrino experiments 
are not very sensitive to the density fluctuations. 
However, the upcoming Borexino experiment \cite{borexino}, 
aiming to detect $^7$Be neutrinos, could 'resolve' this effect.
In order to see that these neutrinos are the ones most 
affected by the fluctuation we have indicated in Fig. 1 
by the vertical lines the position where $^7$Be neutrinos 
fall in for $\delta m^2 \sim 10^{-5}$ eV$^2$. 
\begin{figure}[htb]
\vskip -1.0cm
\centerline{\protect\hbox{
\psfig{file=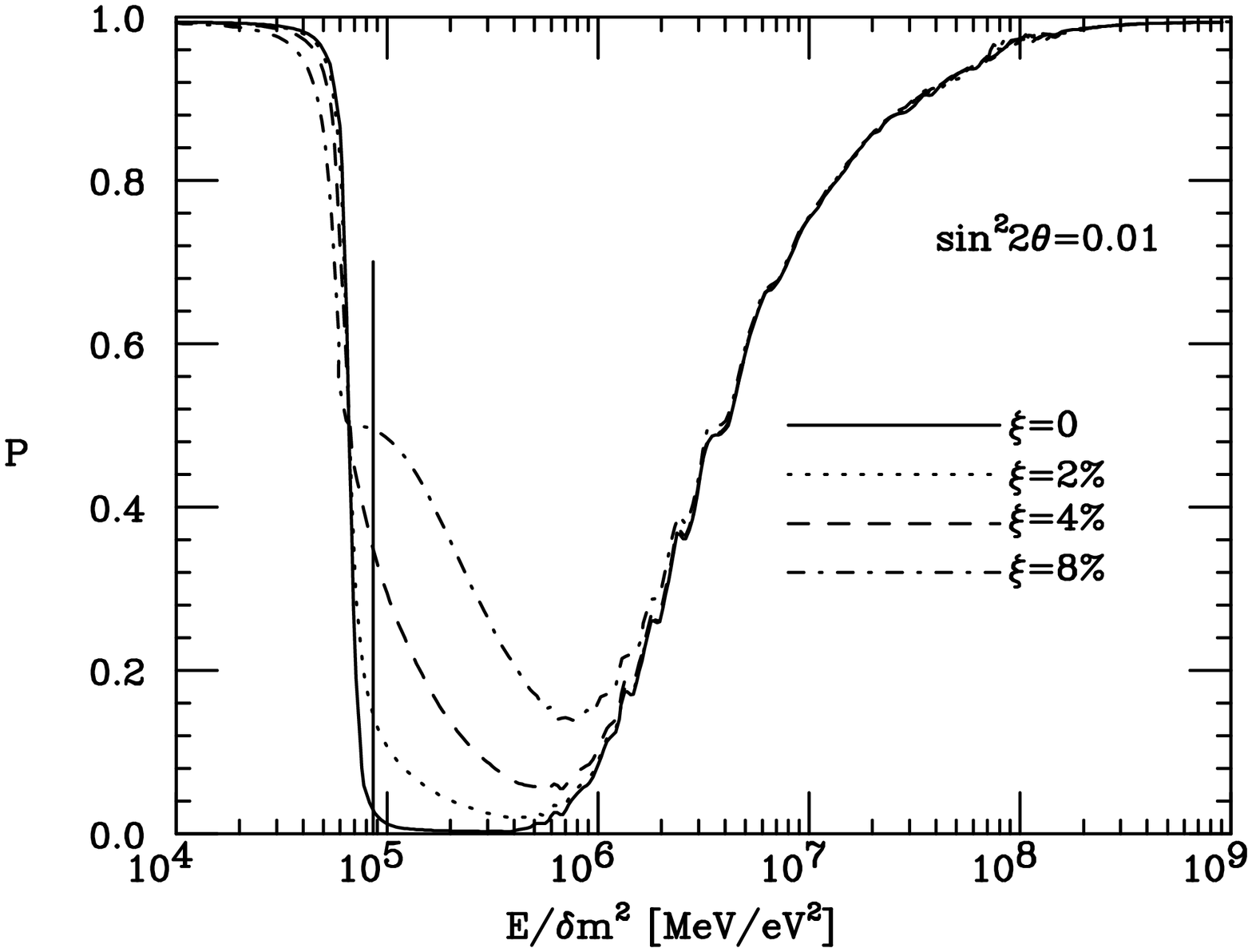,height=5.5cm,width=8.0cm,angle=90}
}}
\vskip -0.8cm
\centerline{\protect\hbox{
\psfig{file=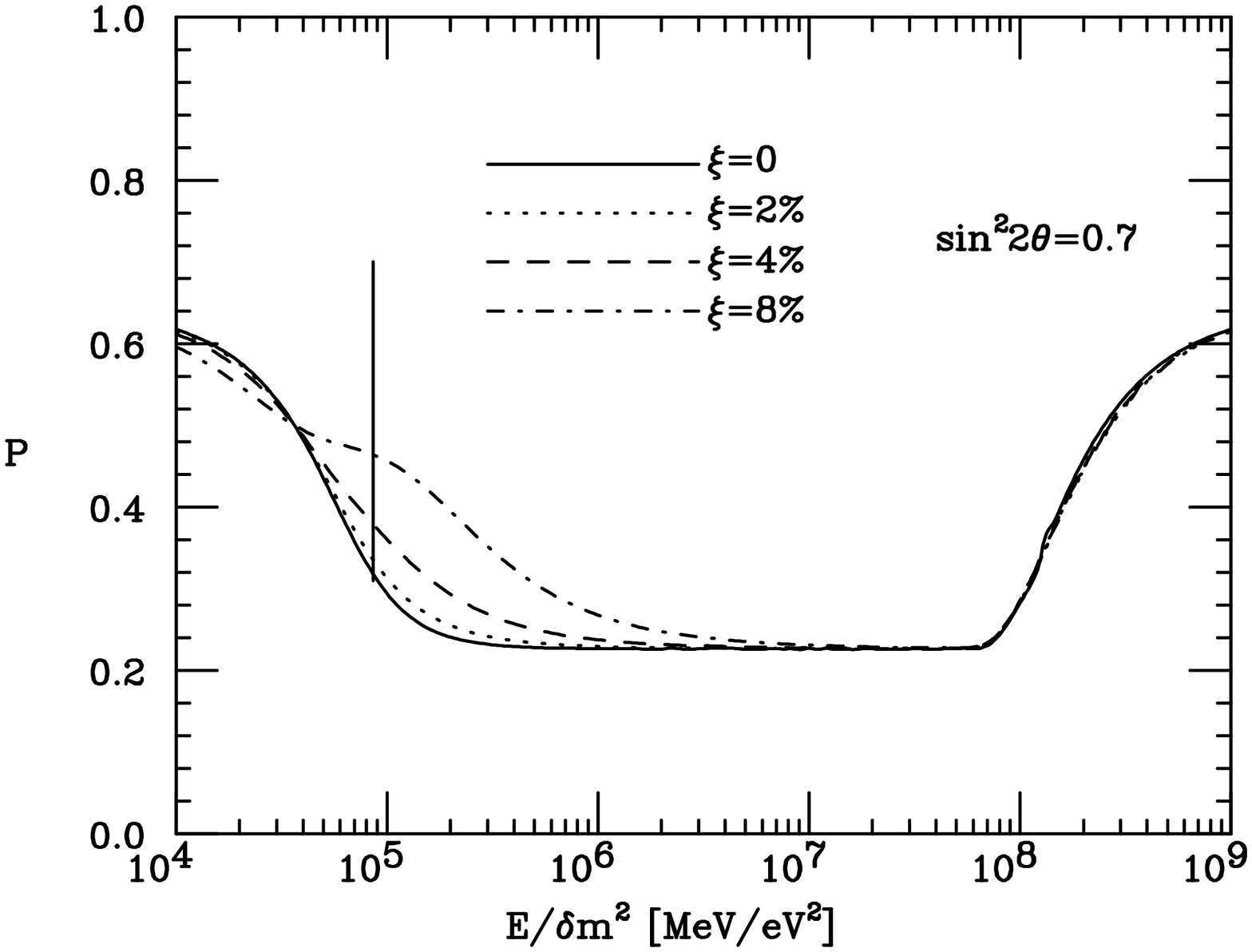,height=5.5cm,width=8.0cm,angle=90}
}}
\vskip -1.5cm
\caption{The averaged solar neutrino survival probability 
as a function of $E/\delta m^2$, which is obtained by 
numerical integration (imposing $L_0 = 0.1 \times \lambda_m$), 
is plotted for $\sin^2 2\theta = 0.01$ (upper panel) and 
0.7 (lower panel). We have used the density profile in 
the Sun from ref. \protect{\cite{BP95}}. 
}
\vskip -0.9cm
\end{figure}
\section{Effect on Borexino Experiment}
Let us see more qualitatively the effect of fluctuation
on Borexino experiment. In Fig. 2 we plot 
the iso $Z_{B_e} \equiv R_{B_e}^{MSW}/R_{B_e}^{BP95}$ contours in the 
Borexino detector for the $\nu_e \to \nu_{\mu,\tau}$ conversion 
without and with density fluctuations. Here 
$R_{B_e}^{MSW}$ and $R_{B_e}^{BP95}$ refer to the expected event 
rate with and without MSW mechanism. 

In the following we assume that 
the standard solar model prediction is correct and 
the solar neutrino deficit is explained by small mixing angle 
MSW solution, either due to $\nu_e \to \nu_{\mu,\tau}$ or 
$\nu_e \to \nu_s$ where $\nu_s$ is a sterile neutrino. 

For the case without fluctuations the Borexino 
signal is expected to be in the range $Z_{Be}\! \sim 0.2\div 0.7$ 
(see the upper panel in Fig. 2)
whereas in the presence of the  fluctuation $\xi= 4\%$ 
(see the lower panel in Fig. 2), the minimal 
allowed value for $Z_{Be}$ becomes higher, $Z_{Be}\!\geq \!0.4$. 
Hence,  if the MSW mechanism is responsible for the 
solar neutrino deficit and Borexino
experiment  detects rather low signal, $Z_{Be}\sim 0.2$
(with good accuracy) this implies that a few \% level of matter 
fluctuations in the central region of the Sun is unlikely. 
The same argument can be applied to $ \nu_e \to \nu_s$ solution
in case 
experiments like Super-Kamiokande and/or SNO
should establish it. 
The expected signal in Borexino is very small $Z_{Be} \approx 0.02$ for 
$\xi =0$ (see the upper panel in Fig. 3). 
On the other hand with $\xi=4\%$, 
the minimum expected signal is 10 times higher than in the
noiseless case, so that if Borexino detects a rate $Z_{Be} \lsim 0.1$ 
(see the lower panel in Fig. 3) this would again exclude noise 
levels above  few \%.

We note that our discussion is valid 
only if $L_0$ is smaller than $\sim 5 \times 10^4 $ 
km which is obtained from the condition (\ref{correlation}) at the 
resonance. If $L_0$ become larger than this value, 
our approach is inadequate 
and moreover, the effect of random  
perturbations ceases to exist \cite{burgess2}. 

\vglue -0.3cm
\begin{figure}[htb]
\vskip -1.0cm
\centerline{\protect\hbox{
\psfig{file=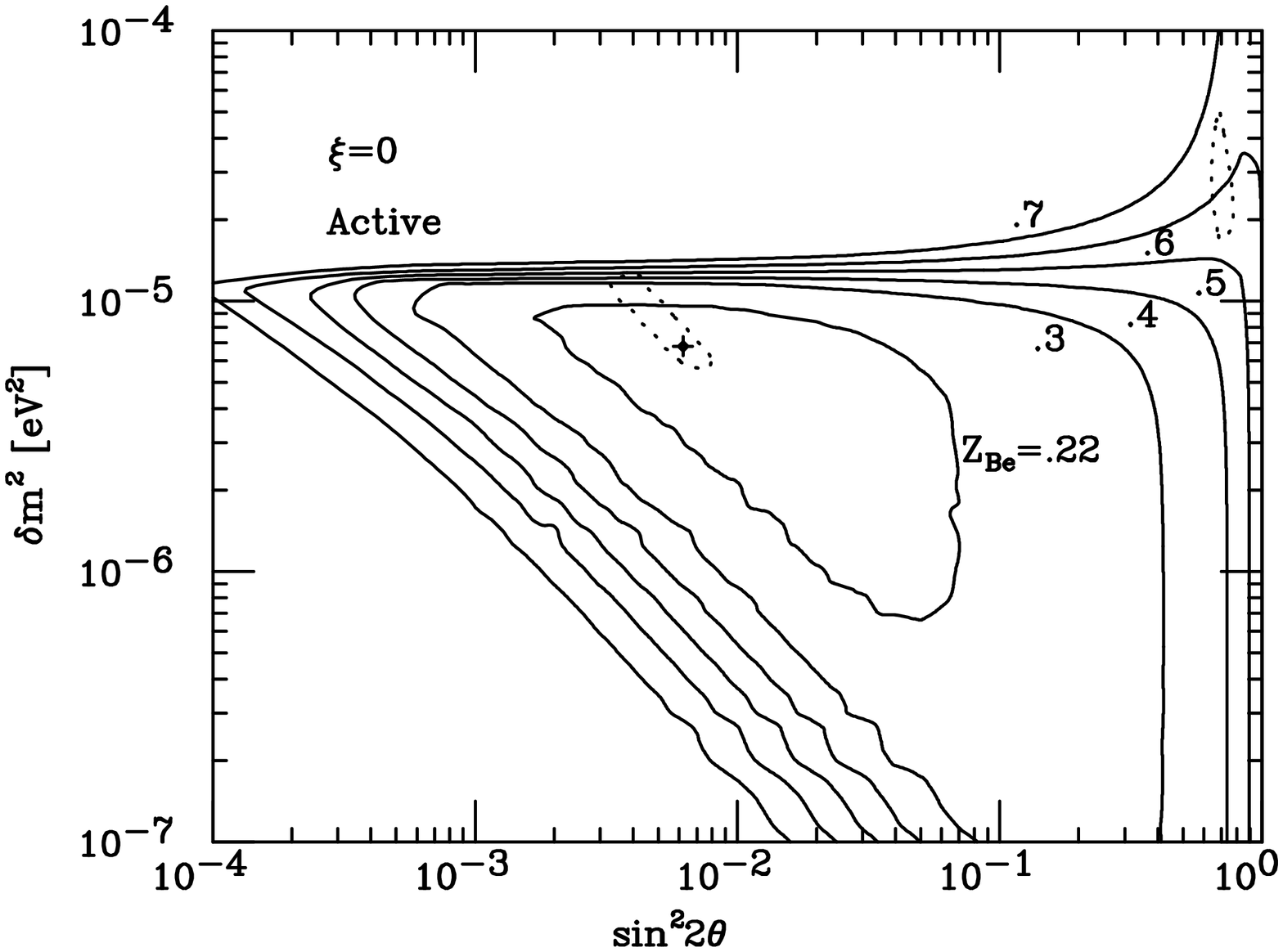,height=5.6cm,width=8.0cm,angle=90}
}}
\vskip -1.0cm
\centerline{\protect\hbox{
\psfig{file=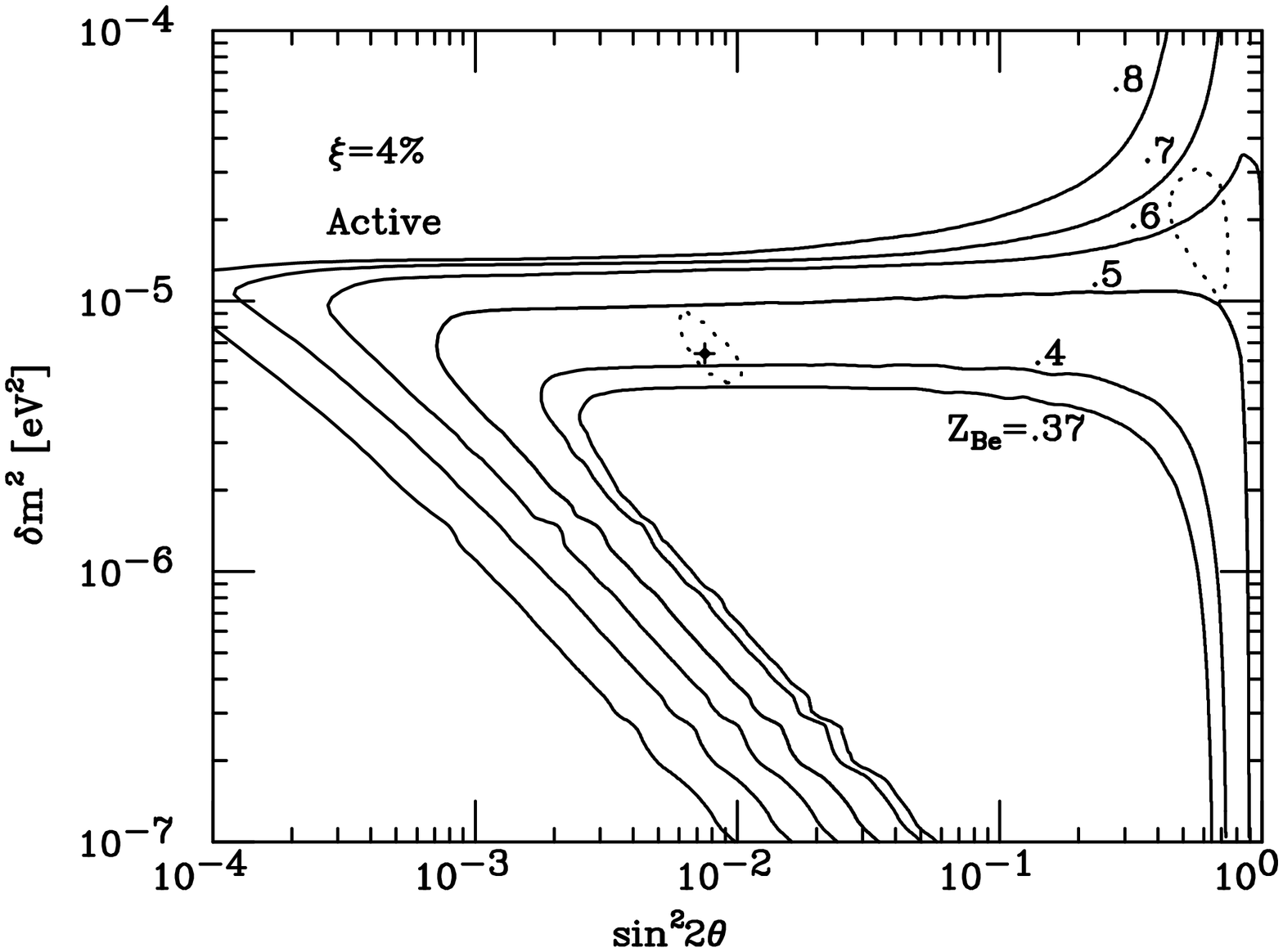,height=5.6cm,width=8.0cm,angle=90}
}}
\vskip -1.5cm
\caption{The iso 
contours of $Z_{B_e}$ (see the text) in the 
Borexino detector for the $\nu_e \to \nu_{\mu,\tau}$ 
conversion are plotted. Upper panel 
refers to the case with no fluctuation, $\xi = 0$ 
whereas the lower panel refers to the case with $\xi = 4$ \%. 
The region allowed at 90\% C.L. is also plotted by 
dotted lines. 
}
\vskip -0.7cm
\end{figure}
\begin{figure}[htb]
\vskip -1.0cm
\centerline{\protect\hbox{
\psfig{file=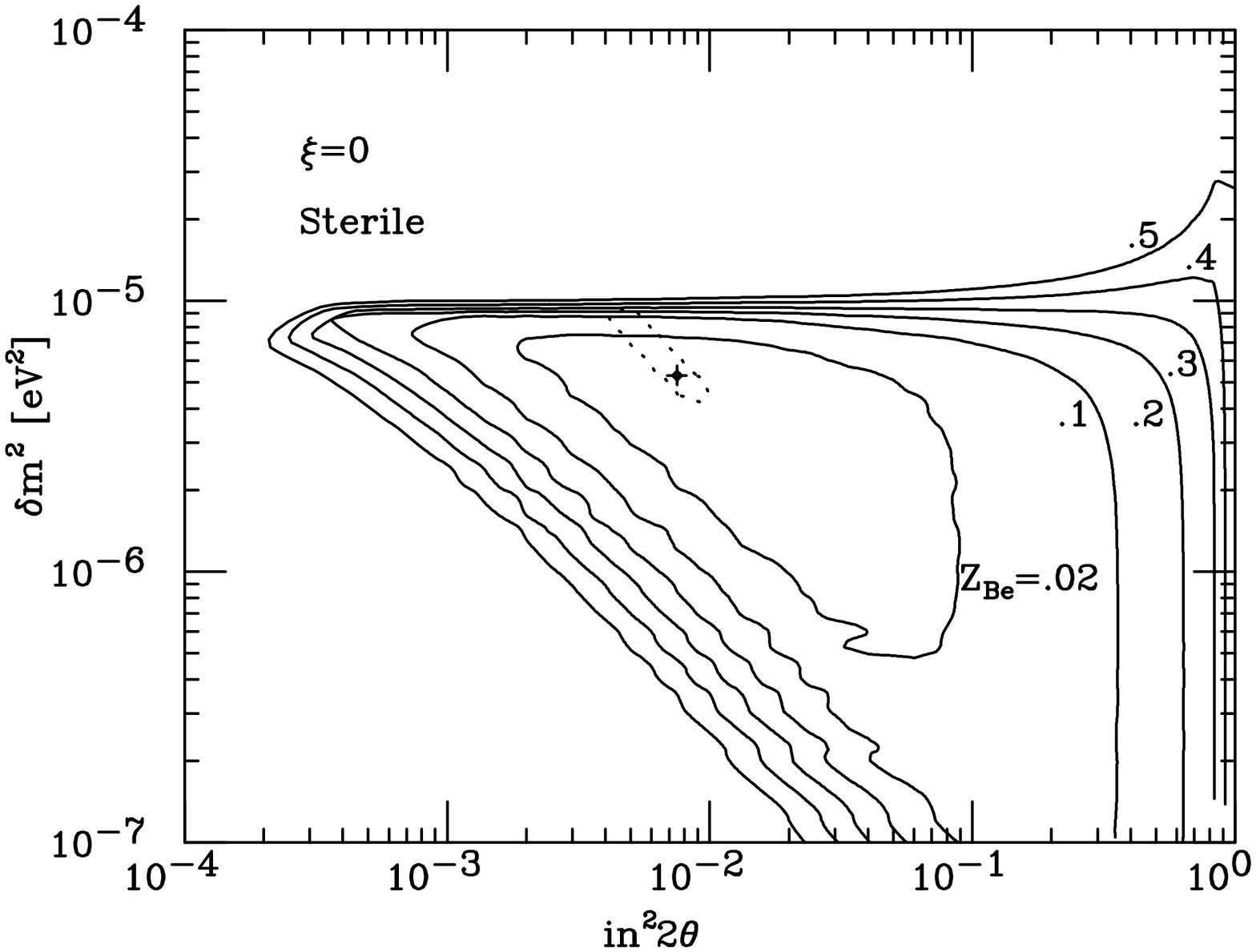,height=5.6cm,width=8.0cm,angle=90}
}}
\vskip -1.0cm
\centerline{\protect\hbox{
\psfig{file=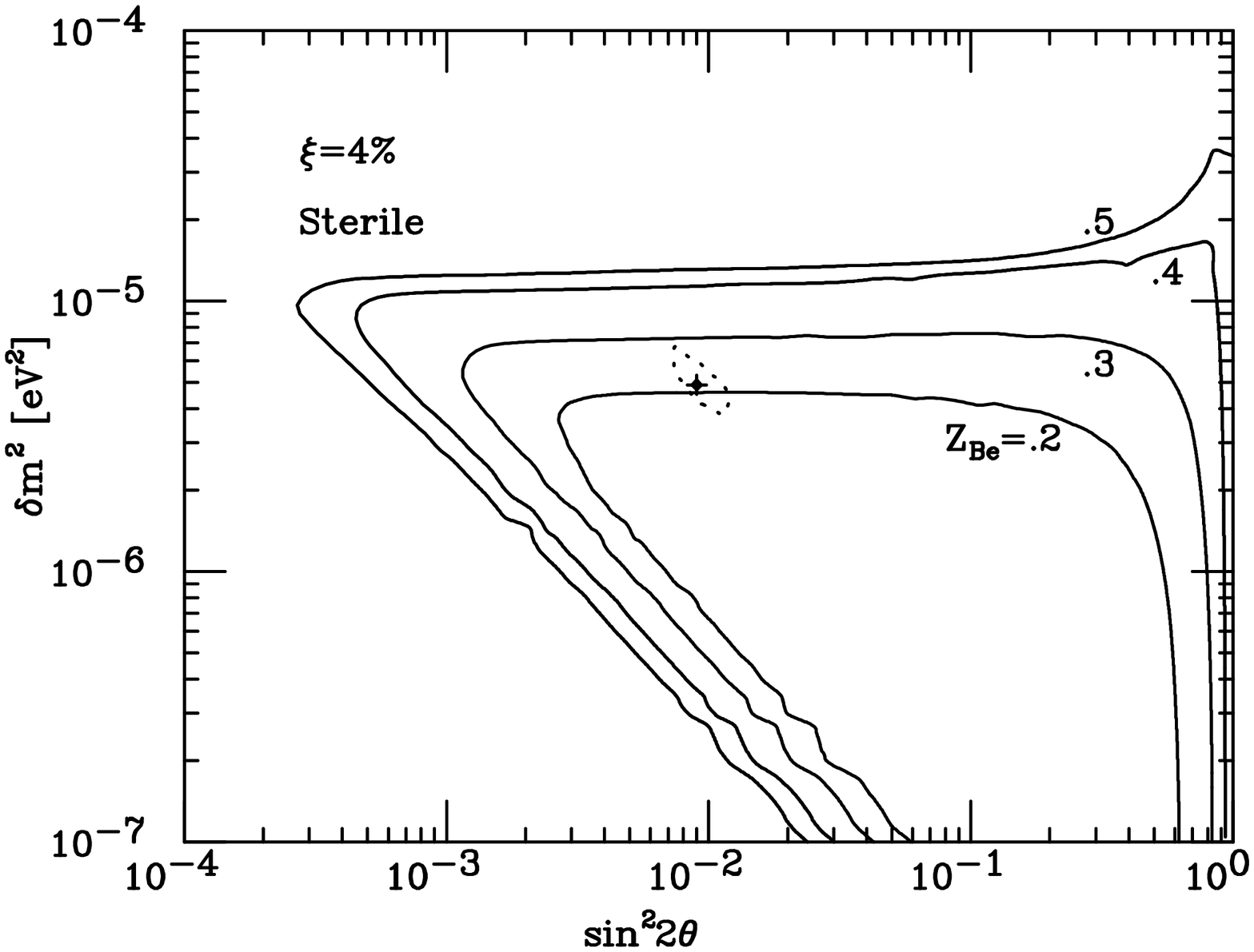,height=5.6cm,width=8.0cm,angle=90}
}}
\vskip -1.5cm
\caption{Same as in Fig. 2 but for the case of 
$\nu_e \to \nu_s$ conversion where $\nu_s$ is a sterile 
state.  
}
\end{figure}
\section{Conclusion}
We have discussed the possibility to test the effect 
of density fluctuation in the Sun by Borexino experiment. 
We conclude that if the small angle MSW solution is 
established and if we see rather small signal in Borexino 
this implies the exclusion of the density fluctuation at 
the level of a few \%, which could be 
compared with independent informations 
from Helioseismology. 

We finally note that ref. \cite{burgess2} concludes that no sizable 
effect from the density fluctuation on the MSW mechanism can be 
expected if the fluctuation is in the form of Helioseismic waves.
We, however, feel that it is too early to conclude that the random 
perturbations at a few \% level with the correlation 
length we considered here is completely unlikely. 

More detailed discussion on this work is found in ref. \cite{noise}. 
\vskip -1.2cm

\end{document}